# Magnetic biophysical characterization of biomimetic polyethylenimine-coated nanoparticles on in vitro silico model

**Enrico Catalano[1]**

[1]  University of Oslo (UiO), Oslo, Norway. email: enrico.catalano@medisin.uio.no
*  Correspondence: Enrico Catalano, e-mail: enrico.catalano@medisin.uio.no

**Abstract:** Understanding the biophysical and magnetic interactions of nanoparticles (NPs) with cell membranes is critical for developing effective nanocarrier systems for drug delivery applications and targeted nanophysics applications. Recent discoveries in nanomedicine can be used to test an *in vitro* system that reproduce a similar tumor model. Iron-oxide nanoparticles can be used for diagnosis, as well as a good carrier of drugs and induced therapeutic for magnetic hyperthermia. In the present study, we engineered polyethylenimine-conjugated superparamagnetic iron-oxide nanoparticles (SPIONs) for the targeted delivery of chemotherapeutics on in vitro silico model. Doxorubicin is used to treat numerous types of tumors including breast cancer. The drug-coated multi-functionalized nanoparticles, were assembled stepwise, with doxorubicin adsorbed to polyethylenimine-iron oxide nanoparticles first, by electrostatic reaction and allowed for the complexation of doxorubicin.. The drug-coated particles were able to inhibit growth and proliferation of resistant cancer cells in vitro, indicating that the system has potential to act as an antimetastatic chemother-motherapeutic agent. Here, we demonstrated a sophisticated strategy to kill in a precise way only cancer cells by conjugating a magnetic nanoparticle to doxorubicin chemotherapeutic.

## 1. Introduction

Nanoparticles (NPs) and nanocarriers, due to their nanometric size, can overcome biological barriers, accumulate preferentially in tumors due to Enhanced permeability and retention (EPR) effect, and specifically target single cancer cells for detection and treatment [1, 2] by minimizing their accumulation in healthy tissues.

Breast cancer is a very common cancer and the most frequent cause for cancer-related deaths in women worldwide, impacting ~2–2.5 million women each year out of 18.08 million new cancer cases (incidence rate of 11.6%) and it was estimated that ~627,000 women died due to breast cancer in 2018. The emergence of nanotechnology offers a paradigm shift in treatment approaches available for breast cancer and triple-negative breast cancer therapy. In the last years improved early diagnostic methods, modern surgery and more effective drugs in the adjuvant setting, increased the average 5-year survival rate for breast cancer to about 90%. However, if distant metastasis occurs, patients are still evaluated as non-curable and have in general only a few years to live while on permanent drug therapy, and current drug therapy is far from optimal, especially in the metastatic setting.

Nanotechnology offers several distinctive pharmacological advantages like nanometric size, longer circulation half-lives, higher drug entrapment, the ability for surface modification, and targeting (active and passive) capabilities. To expand the capacity of targeting delivery of anticancer drugs to tumors, nanoparticles are usually functionalized with targeted antibodies, peptides or other biomolecules. Doxorubicin (DOX) is one of the most effective anticancer drugs used against solid tumors of diverse origins and against breast cancer.

Although doxorubicin is currently considered to be one of the most effective agents in the treatment of breast cancer, the drug use is limited by cardiotoxicity, myelosuppression and palmar plantar erythrodysenthesia and absolute maximum total doses that may be given and resistance to therapy frequently occurs, too. The most diffused molecular mechanism involved in the killing of cancer cells is related to the activation of apoptotic pathway [4].

Nanomedicine products such as Doxil® and Abraxane® have already been extensively used for breast cancer adjuvant therapy with favorable clinical outcomes. However, these products were originally designed for generic anticancer purpose and not specifically for breast cancer treatment. With better nanoengineering and nanotechnology action against breast cancer, a number of novel promising nanotherapeutic strategies and devices can be implemented.

Many different novel cancer therapeutic strategies that have been explored in the last two decades, therapies that are responsive to external physical stimuli coupled with nanoparticles (NPs) that are capable of responding to light, magnetic fields, ultrasound, radio-frequency, or x-ray, have attracted substantial interest from many researchers across the world [17].

Magnetic fields (MF) are non-invasive in nature and have excellent human tissue penetration. Owing to this characteristic of MF, it has been used in clinics for whole body MRI. In recent years, MF has been used to activate drug delivery into the tumor region through field dependent thermal and non-thermal effects. High frequency (>10 kHz) alternating magnetic fields (AMF) responsive MNPs can generate heat by utilizing different physical mechanisms.

Nanosized magnetic nanoparticles are some of the best candidates for the development of more efficient methods for the synthesis of nanodrug delivery vehicles based on iron-oxide nanoparticles.

To this aim, we evaluated the possibility of incorporating DOX into nanoparticles capable of favoring accumulation of the drug into the tumor site. Polyethylenimine (PEI) is extensively known due to its properties and its cationic polymer effect; particularly, branched PEIs of 25-kDa and 750-kDa have proved and these nanoparticles can be associated to doxorubicin because cancer cells frequently become resistant to DOX treatment and some conditions of the tumor microenvironment-such as hypoxia, acidity, defective vasculature, and lymphatic vessels-limit the DOX anticancer effectiveness.

PEI-SPIONs administration and application of a localized magnetic field could thus potentially disrupt tumor vasculature and reduce tumor development. Exploration of the antitumor potency of these nanosystems in combination with other therapies could broaden the spectrum of anticancer agents available. This depicts PEI-SPIONs as four-in-one agents that can deliver: gene transfection, pro-inflammatory stimulation, magnetic targeting, and anti-cancer capacity.

Different anticancer drugs-such as paclitaxel, camptothecin, and tamoxifen-have been efficiently incorporated in nanocarriers, showing an improved antitumor effect [12]. In this study, DOX was loaded in $PEI/Fe_3O_4$ nanoparticles, and the effect of DOX and DOX- $PEI/Fe_3O_4$ was compared in vitro in human breast cancer cell lines, either sensitive or resistant to DOX. Results indicated that use of $PEI/Fe_3O_4$ may be an efficient strategy to deliver DOX in the treatment of breast cancer, since it improves the anti-cancer efficacy and reduces cardiotoxicity. The main issue in the clinical use of doxorubicin is due to its heavy side effects on patients, especially for its intrinsic and cumulative dose-dependent cardiotoxicity [5-7]. To limit these side effects, doxorubicin can be encapsulated with drug nanocarriers that will allow targeted accumulation only on the cancer site and hence limit its dispersion into healthy tissues [8].

## 2. Materials and Methods

### Synthesis of the Fe₃O₄/PEI NPs

The Fe₃O₄/PEI nanoparticles were synthesized using a coprecipitation approach. Briefly, 10 mL of FeCl₂·4H₂O (89 mg) and FeCl₃·6H₂O (157 mg) aqueous solution were mixed with 10 mL of PEI polymer in aqueous solution and the mixture was stirred for 3-5 min to ensure uniform mixing. Afterwards, a sodium hydroxide (1.0 g, 5 mL) aqueous solution was added into the above mixture solution under stirring, and the suspension was continuously stirred for 30 min at 80 °C. Then the mixture was cooled down to room temperature, the product (denoted as Fe₃O₄/PEI NPs) was collected by magnetic separation and purified 5 times with water. Finally, the prepared Fe₃O₄/PEI NPs were redispersed in water.

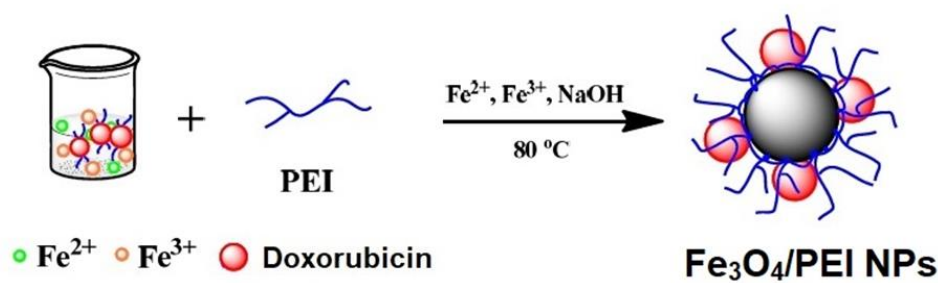

**Figure 1.** Synthesis of polyethylenimine-coated iron-oxide nanoparticles (Fe₃O₄/PEI NPs)

### Conjugation of doxorubicin to Fe₃O₄/PEI NPs

For drug conjugation to modified SPIONs, different concentrations (50 and 100 µg/ml) of Fe₃O₄/PEI NPs were first sonicated with 0.5, 1, 5 and 50 µM concentrations of doxorubicin (DOX – Sigma-Aldrich) solution for 0.5 h and then stirred overnight at room temperature in the dark. DOX loading of modified SPIONs at an initial drug concentration of 50 µM was nearly saturating. All the samples were centrifuged at 18 000 x g for 1 h. The DOX concentration of all the samples was measured using a standard DOX concentration curve, generated with a UV-Vis spectrophotometer (Cary 60 UV-Vis – Agilent) at the wavelength of 233 nm. The drug conjugation to Fe₃O₄/PEI NPs has been illustrated in Figure 2.

### In vitro cell viability of breast cancer cells determination under hyperthermic conditions

Breast cancer cells were exposed to hyperthermia conditions to induce mediated apoptosis. The heating efficiency of the Fe₃O₄/PEI NPs was assessed in the presence of A/C magnetic hyperthermia device. The study was performed at 180 Gauss magnetic field strength with 435 kHz frequency (AMF, conditions: H = 15.4 kA/m, f = 435 kHz) to induce mediated apoptosis of breast cancer cells. Approximately 100 µg/mL of the Fe₃O₄/PEI NPs were dispersed in water and were placed in a 6.5-cm radius copper coil (Greiner BioOne, Frickenhausen, Germany). The sensitivity of MCF-7 and MDA-MB-231 cells to hyperthermic temperatures and the effect of the MNP formulations on cell viability in the presence and absence of heat were assessed. At 24 h after seeding 5000 cells/well in a 96-well plate, cells were incubated with either medium, DOX-Fe₃O₄/PEI NPs in a concentration of 100 µg Fe/ml or the equivalent molar amount of free DOX for 24 h at 37°C.

### Physicochemical characterization of nanoparticles

X-Ray powder diffraction (XRD) measurements were performed on a powder sample of the $Fe_3O_4$ nanoparticles using a Rigaku D/Ultima IV X-ray diffractometer (Rigaku, JAPAN), which was operated at 35 kV and 40 mA at a scan rate of 0.4 deg $s^{-1}$ and $2\theta$ ranges from 10° to 90° at room temperature using Cu-Ka radiation ($\lambda = 0.1542$ nm). Transmission electron microscope (TEM) images were obtained at 200 kV by a JEM 2010 (JEOL, JAPAN) instrument. A drop of suspension of magnetite in ethanol (20 mg $mL^{-1}$) before and after modification was placed on a carbon-coated copper grid (Xinxin Bairui Corp, Beijing, China) and air dried. Fourier transform infrared (FTIR) spectra were recorded on a NEXLIS FTIR (Nicolet Corporation) and samples were dried at 90 C vacuum for at least 3 h prior to fabrication of the KBr pellet. In this context 4 mg of each sample was thoroughly mixed and crushed with 400 mg of KBr, and 70 mg of that mixture was used for pellet fabrication. Thirty-two scans of the region between 400 and 4000 $cm^{-1}$ were collected for each FTIR spectrum recorded.

### Dynamic light scattering (DLS) measurement of particle size distribution

DLS was applied to measure size distribution of released DOX-SPIONs. Briefly, DOX-SPIONs nanoparticles were suspended in 200 μl of PBS (pH 7.4 or 6.6), followed by shaking at 300 rpm in a thermomixer (Eppendorf, Germany) at 37°C for 12 h. Supernatants containing released DOX-SPIONs nanoparticles were collected by centrifugation at 2,000 rpm for 4 min. SPIONs controls were prepared by the same procedure. DOX-SPIONs controls were prepared by direct dispersion of 5 μl DOX DMF solution into 2 mL PBS (pH 7.4 or 6.6). Particle size was measured with a Zetasizer Nano-ZS (Malvern Instruments Ltd, Worcestershire, UK) with measurements made using intensity average.

### *In vitro* cellular uptake assay

ICP-OES was carried out to investigate the cellular uptake of the $Fe_3O_4$/PEI NPs by MCF-7 or MDA-MB-231 cells. Both kinds of cells were seeded into 12-well plates at a density of $2 \times 10^5$ cells/well. After incubation at 37 °C and 5% CO2 overnight, the medium was replaced with fresh medium containing the $Fe_3O_4$/PEI NPs at different Fe concentrations (0.2, 0.4 and 0.6 mM, respectively). After 4 h incubation, the medium was discarded carefully and the cells were washed with PBS for 5 times, trypsinized, centrifuged, and counted by Handheld Automated Cell Counter (Millipore, Billerica, MA). The remaining cells were lysed using an aqua regia solution (1.0 mL) for 2 days, and then the Fe uptake in the cells was quantified by ICP-OES.

### Cell cultures

The human breast cancer cell lines MCF-7 (ATCC - HTB-22) and MDA-MB-231 (ATCC - HTB-26) cells were obtained from the American Type Culture Collection and cultured in DMEM, containing 1% (v/v) penicillin-streptomycin and 10% (v/v) heat-inactivated FBS, respectively. FBS and penicillin-streptomycin were obtained from GIBCO - Life Technologies, Carlsbad, CA, USA.

Human mammary epithelial cells (MCF 10A, ATCC, CRL-10317) was cultivated in ™ Mammary Epithelial Cell Growth Medium (MEGM, Lonza/Clonetics) fortified with 2 mM L-glutamine, 100 ng/ml cholera toxin, 10% fetal bovine serum (FBS, Gibco, Thermo-Fisher, Pittsburg, PA, US) and 1% antibiotics/antimycotics (penicillin/streptomycin/gentamycin, Gibco) (complete medium). Cells were incubated in a humidified 5% $CO_2$ atmosphere at 37°C and passaged twice weekly at a 1:4-5 concentration.

**Cytotoxic Assay**

For toxicity assays, cells were treated with PEI-SPIONs (10-100 µg/ml) for 24 h. PEI-SPIONs concentrations refer to iron amounts determined by inductively coupled plasma atomic emission spectroscopy. Direct cytotoxicity of SPIONs on breast cancer cells was evaluated following the ISO standard 10993-5:2009 on Biological Evaluation of Medical Devices instructions [3]. Cells were seeded in 24-well plates (1.6 x $10^4$ cells per well) in complete DMEM medium and incubated for 24, 48 and 72 h at 37°C in 5% $CO_2$. Different combinations of hyperthermia and targeted drug delivery of doxorubicin were tested on breast cancer cells to observe their efficacy. Fresh medium without MNPs was used as control. Cell viability was evaluated after 24, 48 and 72 hours by the (3-(4,5-Dimethylthi-azol-2-yl)-2,5-diphenyltetrazolium bromide colorimetric assay (MTT, Sigma). Briefly, 100 µl of MTT solution (3 mg/ml in PBS) were added to each specimen and incubated for 4 hours in the dark at 37°C; formazan crystals were then dissolved in 100 µl of dimethyl sulphoxyhde (DMSO, Sigma- Aldrich) and 50 µl were collected and centrifuged to remove any debris. Supernatant optical density (o.d.) was evaluated at 570 nm. The mean optical densities obtained from control specimens were taken as 100% viability. Cell viability was calculated as follow: (experimental o.d. / mean control o.d.)*100. Experiments were performed with four replicates at each experimental time.

**Statistical Analysis**

All experimental data derived from DLS analysis and LDH assay were statistically analyzed with Kruskal–Wallis one-way ANOVA and Dunn's post hoc tests (GraphPad Prism 8) to analyze significant differences between groups. The significance level was set at 5%.

## 3. Results

**Physicochemical Characterization**

The PEI were then used as a stabilizer to prepare $Fe_3O_4$/PEI NPs via a co-precipitation route. UV-vis spectroscopy was used to characterize the products (Fig. 2a). $Fe_3O_4$/PEI NPs display a surface plasmon resonance (SPR) peak at 545 nm, while the $Fe_3O_4$/PEI NPs synthesized using PEI as a stabilizer under similar conditions do not exhibit such a peak in the same region.

X-ray diffractometry of magnetic nanoparticles was used to identify the crystal structure and estimate the crystallite size of as-prepared nanoparticles (Fig. 2b). The lattice constant *a* was measured to be 8.310 Å, which was compared with the lattice parameter for the magnetite of 8.39 Å [3]. As shown in Fig. 1b, the peaks indexed as planes (220), (311), (400), (422), (511) and (440) correspond to a cubic unit cell, characteristic of a cubic spinel structure [4]. In addition, the strongest reflection from the (311) plane is also the characteristic of such phase. Therefore, it is confirmed that the crystalline structure of obtained magnetite nanoparticles is an inverse spinel type oxide, in accord with that of standard data (JCPDS 72-2303).

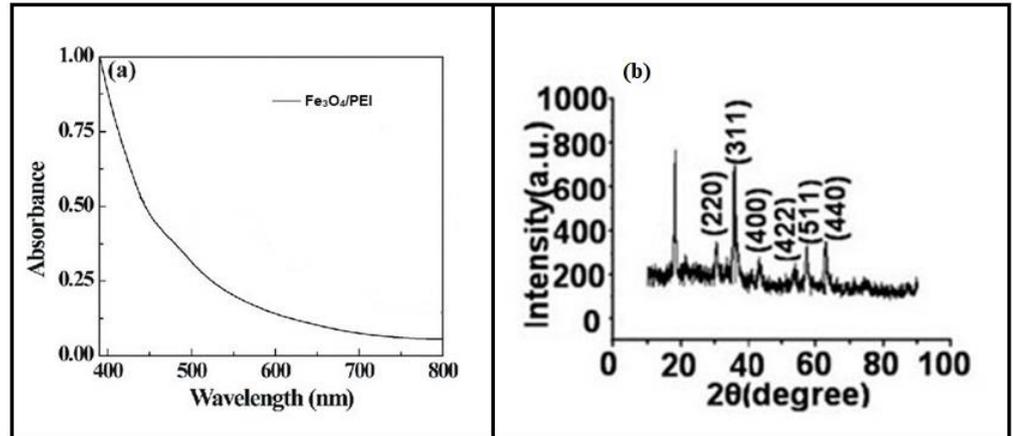

**Figure 2.** (a) UV-vis spectra of the Fe₃O₄/PEI NPs; (b) XRD pattern of the Fe₃O₄/PEI NPs.

**Table 1.** Zeta potential, hydrodynamic size, and polydispersity index of the Fe₃O₄/PEI NPs and DOX-Fe₃O₄/PEI NPs. Data are provided as mean ± S.D. (n= 3)

| Materials | Zeta potential (mV) | Hydrodynamic size (nm) | Polydispersity index (PDI) |
|---|---|---|---|
| **Fe₃O₄/PEI NPs** | +38.5 ± 1.3 | 95.6 ± 4.3 | 0.56 ± 0.08 |
| **DOX-Fe₃O₄/PEI NPs** | +33.1 ± 1.5 | 137.2 ± 5.6 | 0.62 ± 0.05 |

**TEM images**

The morphology and structure of the Fe₃O₄/PEI NPs were characterized by TEM. The TEM image of the Fe₃O₄/PEI NPs is not very clear, which could be due to the existence of a large amount of macromolecular coating of PEI on the particle surface. Some aggregated or interconnected particles in the TEM image could be ascribed to the TEM sample preparation process, especially the air-drying process of the aqueous suspension. It is important to note that the non-uniform distribution of crystals around the surface of the Fe₃O₄ NPs is beneficial for the NPs to have non-compromised T2 relaxivity, because the accessibility of water protons does not have appreciable changes when compared to the Fe₃O₄ NPs without PEI coating. The EDS spectrum further confirms the presence of Fe elements in the Fe₃O₄/PEI NPs (Fig. 2). The existence of Cu element should be due to the copper grid used for TEM sample preparation.

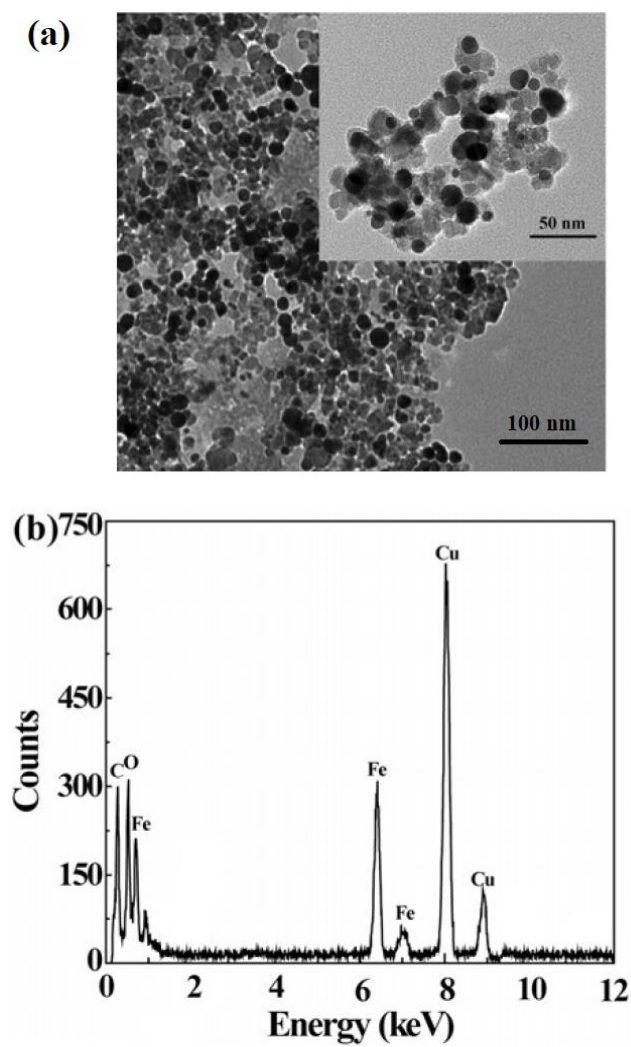

**Figure 3.** TEM image (a) and EDS spectrum (b) of the Fe₃O₄/PEI NPs. Inset of (a) shows the enlarged TEM image of the NPs.

**Colloidal stability and hydrodynamic behaviour of SPIONs**

The colloidal stability of Fe₃O₄/PEI NPs was investigated before cell viability evaluation. The hydrodynamic size was measured in a narrow range by DLS evaluation after 60 days, as shown in Fig. 5. There was no sign of aggregation and flocculation or subsidence after 60 days of storage (Fig. 5).

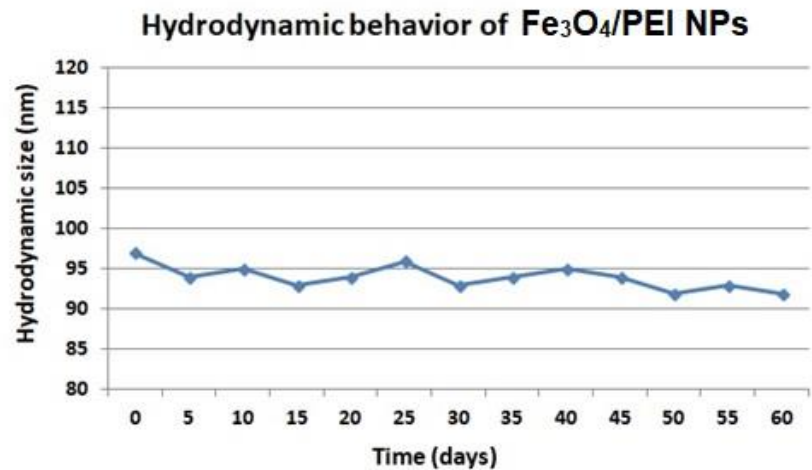

**Figure 4.** Hydrodynamic sizes of Fe₃O₄/PEI NPs in aqueous solution as a function of storage time.

### Internalization of SPIONs into cells

The results of ICP-MS showed that higher amount of iron (45.8 ± 1.8 pg) was taken up in higher SPIONs concentration (100 µg/ml) (concentration-dependent). After identical sample preparation in the control, 10.1 ± 0.9 pg iron per cell could be detected. In the case of 50, 100 µg/ml of SPIONs, the numbers were 24.9 and 45.8 pg/cell, respectively for Fe₃O₄/PEI NPs (Fig. 6).

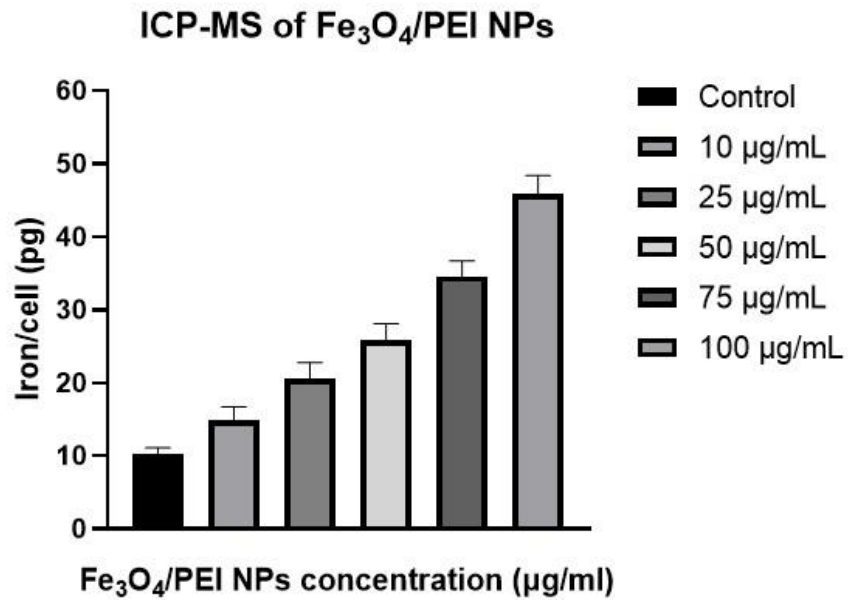

**Figure 5.** ICP-MS of different concentrations of Fe₃O₄/PEI NPs.

### Material toxicity testing of magnetic nanoparticles leach solution

The values of cell viability revealed no significant difference in MCF-10A cell growth compared with that in the control group after Fe₃O₄/PEI NPs nanoparticle leaching solution (100%, 75%, 50%, and 25%) was added. The relative growth rate is shown in Table 2. The results indicate that the 100% leaching solution of Fe₃O₄/PEI NPs is cytocompatible toward MCF-10A cells. Considering the relative growth rate, the eluates solutions of Fe₃O₄/PEI NPs showed a cell viability on MCF-10A cells comparable to control.

| Fe₃O₄/PEI nanoparticles | | |
|---|---|---|
| **Groups** | **OD** | **Relative growth rate (%)** |
| **Control** | 0.289 + 0.005 | 100 |
| **25% extract liquid** | 0,277 + 0.004 | 95.9 |
| **50% extract liquid** | 0.271 + 0.006 | 93.8 |
| **75% extract liquid** | 0.261 + 0.008 | 90.6 |
| **100% extract liquid** | 0.254 + 0.006 | 87.8 |

**Table 2.** Leaching extract solutions for Fe₃O₄/PEI NPs nanoparticles.

**Direct contact cytotoxicity**

The same conditions were verified on MCF-10A cells. For MCF-10A cells the cell viability after 24 hours (MTT assay) range between 80% and 88% for Fe₃O₄/PEI NPs (Fig. 7). DMSO (10% v/v) was used like negative control. The cell viability after 48 hours (MTT assay) range between 83% and 91% for Fe₃O₄/PEI NPs (Fig. 7). The cell viability after 72 hours (MTT assay) range between 82% and 92% for Fe₃O₄/PEI NPs (Fig. 7). Fe₃O₄/PEI NPs did not affect the viability of mammary epithelial cells in all the experimental conditions. These data of cell viability showed cytocompatibility and stability of the layer of polyethylenimine comparable to control with a selectivity for cancer cells compared to non-cancerous cells.

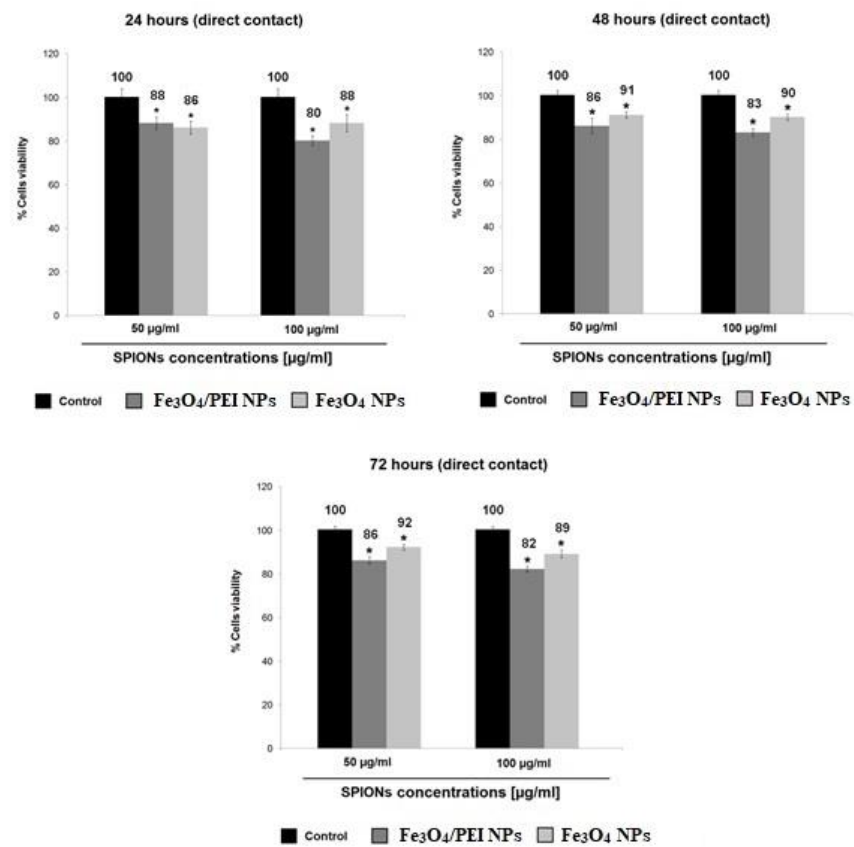

**Figure 6.** Direct contact cytotoxicity evaluation of Fe₃O₄ NPs and Fe₃O₄/PEI NPs (50 and 100 μg/mL) on MCF-10A cells at different time points. *P < 0.05 compared with control samples.

**Direct cytotoxicity evaluation of SPIONs and DOX-SPIONs on cancer cells**

The cytotoxicity of DOX loaded Fe₃O₄/PEI NPs was compared with free drug DOX in the absence and presence of magnetic field, also the probable cytotoxicity of the targeted and non-targeted blank nanoparticles was checked. Cell viability assay showed that in MCF-7 and MDA-MB-231 cells the cell survival percentage was decreased significantly (p<0.05) in targeted nanoparticles group compared to free DOX and naked magnetic nanoparticles in most drug concentrations both in absence and presence of magnetic field without toxic effects on normal mammary epithelial cells (MCF-10A). These data indicate that DOX-Fe₃O₄/PEI NPs and the presence of magnetic hyperthermia led to increasing cytotoxic effects of the nanoparticles on breast cancer cells.

A selective efficacy was observed for DOX-SPIONs on breast cancer cells. The results demonstrated the potential of the DOX-Fe₃O₄/PEI NPs to achieve dual tumor targeting by magnetic field-guided in breast cancer cells and exploit the incredible possibilities to kill in a target way the cancer cells.

Drug-loading efficiency of doxorubicin showed better results for the Fe₃O₄/PEI NPs as compared to the bare ones. For hyperthermia treatment, cells were exposed to hyperthermic conditions for 30 minutes, corresponding to a temperature dosage of 90 cumulative equivalent minutes, or else cells will be left in the incubator at 37°C. At 24, 28, 72 h after hyperthermia, cells viability was tested. The in vitro efficacy was also tested without hyperthermia treatment. The incubation of MCF-7 and MDA-MB-231 human breast cancer cells with doxorubicin-loaded and doxorubicin-loaded polyethylenimine -coated SPIONs, for 24, 48, 72 h, showed significant $IC_{50}$ and $IC_{90}$, respectively, after 72 h of incubation (Figure 8 and 9). While 90% and 93% growth inhibition were seen in MCF-7 and MDA-MB-231 cells after the 72-h exposure to the doxorubicin Fe₃O₄/PEI NPs (p < 0.05), any type of cytotoxic effect was observed on control MCF-10A cells (ATCC, mammary epithelial cells) (Figure 10). These data indicate that DOX-n(COOH)-SPIONs and the presence of magnetic hyperthermia led to increasing cytotoxic effects of the nanoparticles on breast cancer cells.

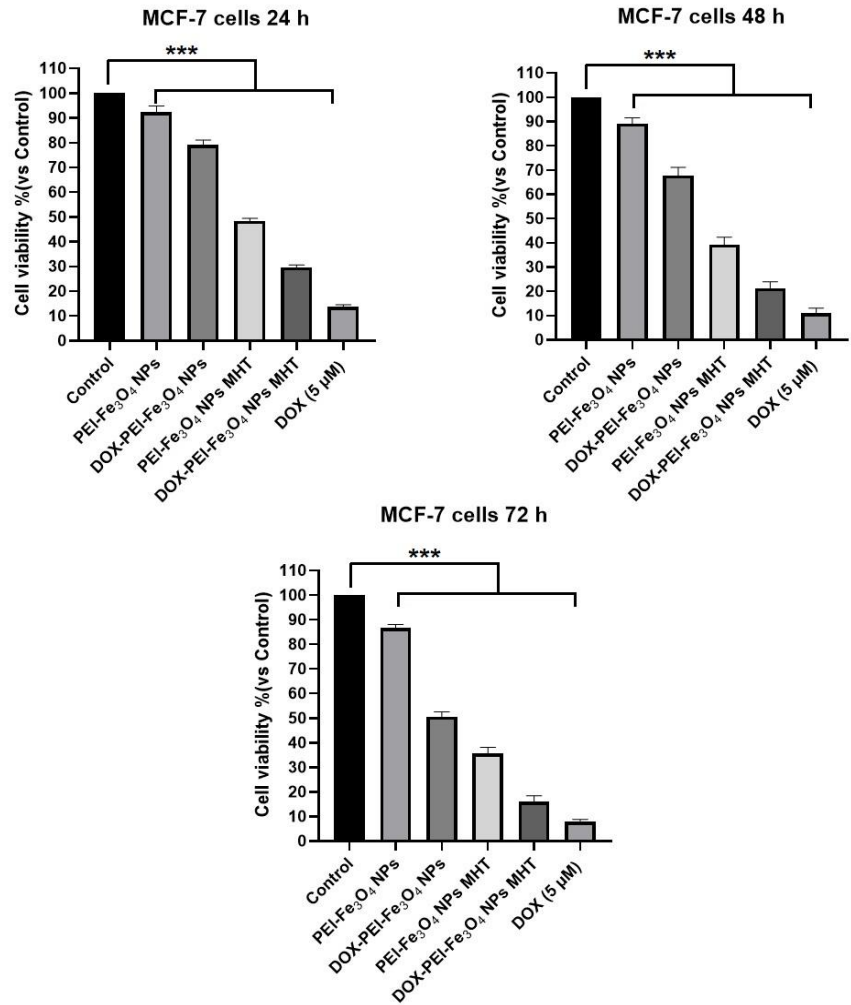

**Figure 7.** Direct contact cytotoxicity evaluation of Fe₃O₄/PEI NPs (50 μg/mL) conjugated to Doxorubicin (5 μM) (DOX) and with induced magnetic hyperthermia using breast cancer cells (MCF-7 cells) at different time points. Data are shown as the mean ± standard error of the mean (n = 4). ***P < 0.001 compared with control samples.

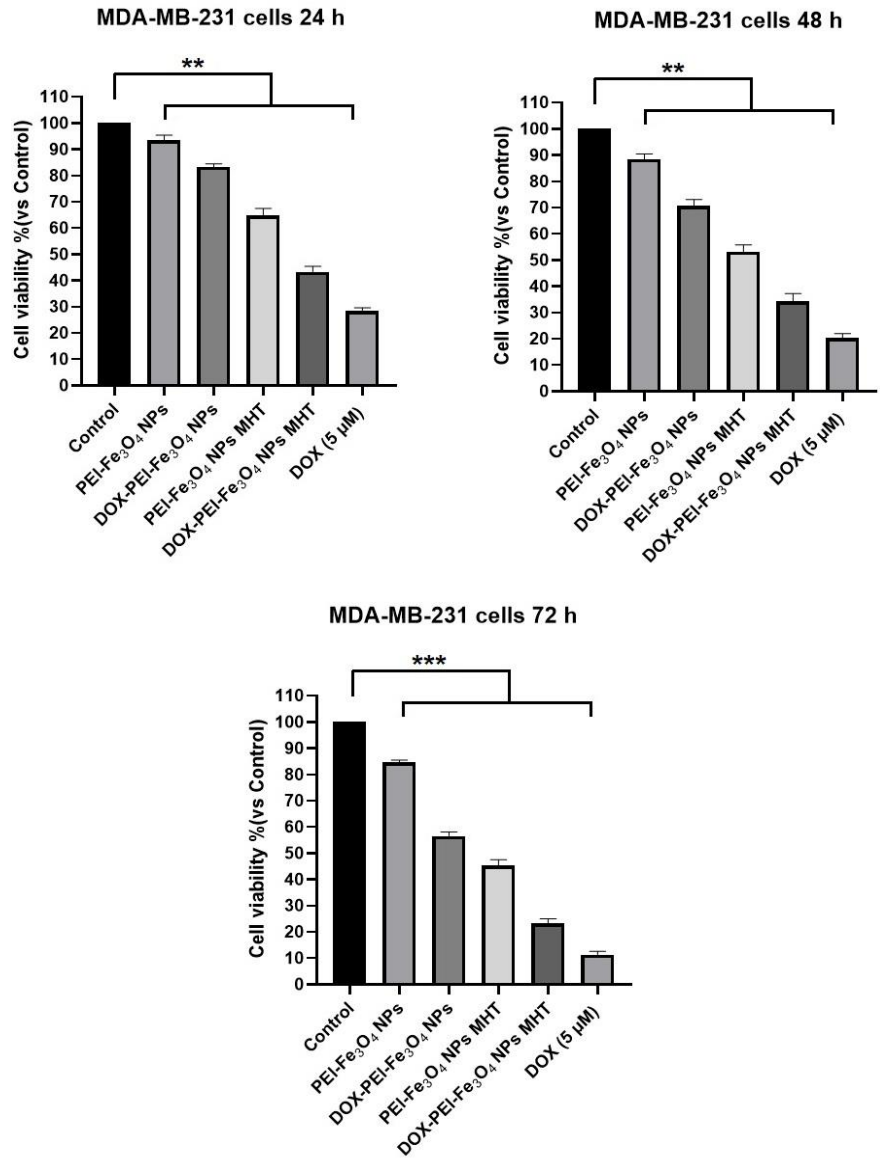

**Figure 8.** Direct contact cytotoxicity evaluation of Fe$_3$O$_4$/PEI NPs (50 μg/mL) conjugated to Doxorubicin (5 μM) (DOX) and with induced magnetic hyperthermia using breast cancer cells (MDA-MB-231 cells) at different time points. Data are shown as the mean ± standard error of the mean (n = 4). *P < 0.05 compared with control samples. **P < 0.05 compared to control samples and DOX 5 μM.

### Results of toxicity in 3D structure on breast cancer cells

3D spheroid cultures, constitute a more physiologically model than 2D cell cultures for the evaluation of novel therapeutic strategies. As observed for 2D cell cultures, the invasion capacity of matrigel-embedded 3D cultures of MCF-7 and MDA-MB-231 cells was significantly reduced after controlled nanodelivery release of doxorubicin with reverse flexible ionic bond for the release of doxorubicin from iron oxide nanoparticles.

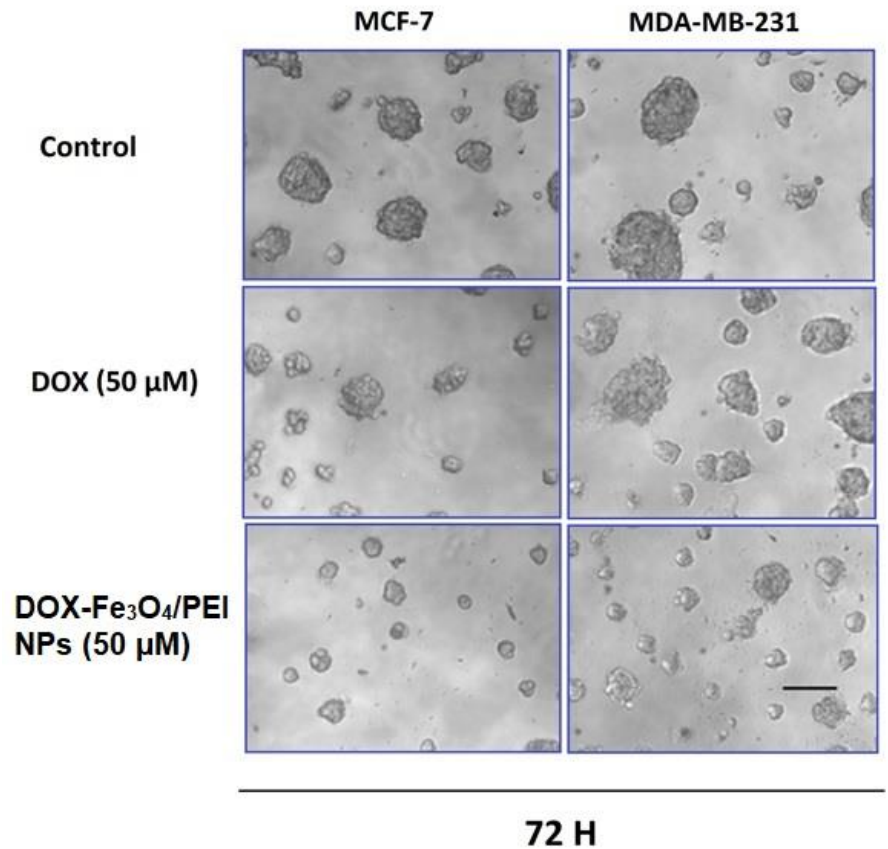

**Figure 9.** Invasion capacity of matrigel-embedded 3D cultures of MCF-7 and MDA-MB-231 cells is reduced with DOX-Fe$_3$O$_4$/PEI NPs. Cells were grown in a semi-solid matrigel matrix. Then, 3D cultures were exposed to the indicated doses of the drugs. After 72 h was taken pictures and quantified the spheres size. Scale bar= 100 μm.

**Apoptosis evaluation**

The results revealed a dose-dependent induction of early or late apoptotic cell death in the breast cancer cell lines (Figure 11). Compared to MDA-MB-231 cells, the MCF-7 cells showed more cell death. MCF-7 cells showed 91% apoptosis after the administration of DOX-loaded Fe$_3$O$_4$/PEI NPs (5 μM; p < 0.05). While, 87.1% apoptosis was seen in the MDA-MB-231 cells under same conditions (p < 0.05), which indicates that polyethileimmine-fuctionalized SPIONs could enhance the DOX-induced apoptosis in both of the human breast cancer cell lines. For the cell death onset analysis, MCF-7 and MDA-MB-231 cells were treated separately with DOX, SPIONs, SPIONs MHT and DOX- Fe$_3$O$_4$/PEI NPs AMF for 24 h of time. Apoptotic cell death was analyzed by staining the cells with Annexin-V-FLUOS staining kit and analyzed by flow cytometry. The results revealed a dose-dependent induction of early or late apoptotic cell death in these two cell lines (Figure 11). Compared to MDA-MB-231 cells, the MCF-7 cells showed more cell death. MCF-7 cells showed 91% apoptosis after the administration of DOX-loaded SPIONs (5 μM; p < 0.05). While 87.1% apoptosis was seen in MDA-MB-231 cells under same conditions (p < 0.05), which indicates that hydroxyl-modified SPIONs could enhance the DOX-induced apoptosis in both human breast cancer cell lines.

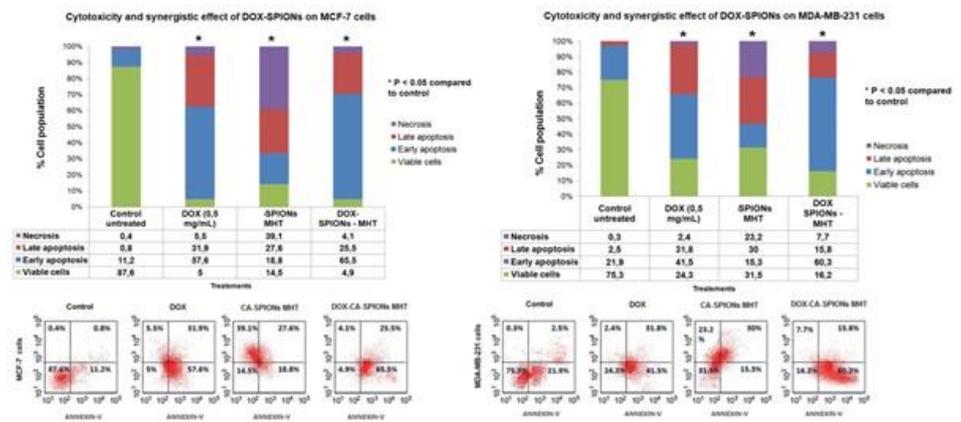

**Figure 10.** Doxorubicin (DOX)-loaded Fe₃O₄/PEI NPs system induces apoptosis in MCF-7 and MDA-MB-231 cells. Both cancer cell lines were treated separately with DOX, SPIONs, Fe₃O₄/PEI NPs, and DOX-Fe₃O₄/PEI NPs for 24 h. Apoptotic cell death was detected by staining the cells with Annexin-V/PI kit and analyzed by flow cytometry.

### In vitro toxicity of doxorubicin loading onto the SPIONs

$IC_{50}$ and $IC_{90}$ analyses of MCF-7 and MDA-MB-231 cell lines, after 24, 48, 72, 96 h exposure to DOX-loaded Fe₃O₄/PEI NPs were recorded at $1.9 \pm 0.5$ and $4.2 \pm 0.7$ mM concentration of drug, while $IC_{90}$ of both cell lines was seen at $4.3 \pm 0.7$ and $6.8 \pm 1.0$ mM DOX concentration after 96 h for MCF-7 and MDA-MB-231 cells respectively. This demonstrates that the MDA-MB-231 cells were more resistant to the lower concentrations of drug as compared to the MCF-7 cells (Table 3; $p < 0.05$).

**Table 3.** $IC_{50}$ and $IC_{90}$ of MCF-7 and MDA-MB-231 cell lines after 24, 48, 72, 96 exposure to DOX- Fe₃O₄/PEI NPs

| | $IC_{50}$ for DOX-loaded Fe₃O₄/PEI NPs (mM) | | | |
|---|---|---|---|---|
| **Type of cell line** | | | | |
| **MCF-7** | $4.0 \pm 0.9$ | $3.1 \pm 0.8$ | $2.4 \pm 0.7$ | $1.9 \pm 0.5$ |
| **MDA-MB-231** | $6.4 \pm 1.8$ | $5.3 \pm 1.2$ | $4.8 \pm 0.9$ | $4.2 \pm 0.7$ |
| | | | | |
| | $IC_{90}$ for DOX-loaded Fe₃O₄/PEI NPs (mM) | | | |
| **Type of cell line** | | | | |
| **MCF-7** | $8.9 \pm 1.5$ | $7.6 \pm 1.0$ | $6.1 \pm 0.6$ | $4.3 \pm 0.7$ |
| **MDA-MB-231** | $10.6 \pm 3.4$ | $9.4 \pm 2.1$ | $7.3 \pm 0.8$ | $6.8 \pm 1.0$ |

DOX, doxorubicin. The $IC_{50}$ and $IC_{90}$ are defined as the concentrations causing 50% and 90% growth inhibition in treated cells, respectively, when compared to control cells. Values are means ± SEM of at least three separate experiments.

### DOX release profile

The DOX release profiles of $Fe_3O_4$/PEI NPs were analyzed to check the free and the conjugated drug, at the pH range of 1.5-7.0. At Ph 7.0, a small amount of the drug release was observed after the incubation period of 48 h.

The DOX release profiles of the free and the conjugated drug, at the pH range of 1.5-7.0. At pH 7.0, a small amount of the drug release was observed after the incubation period of 48 h. This is a desirable characteristic as the pH 7.4 is the undesired pH for the proper release of the drugs from the nanoconjugated drug carrier. This will also prevent the premature release of the drugs before the nanoconjugates reach the cancer cells. It shows that the pH 6.0 provided the desirable conditions for the proper drug release. The first 10 h represent the period of initial rapid release, followed by a steady state.

This pH-dependent drug release behavior is favorable for the chemotherapeutic process as it can significantly reduce the preterm drug release on the body pH level (pH 7.4) and maximizing the amount of drug reaching the target tumor cells, once the drug-loaded magnetic nanoparticles internalize and enter the tumor by endocytosis (pH 4.5-6.5) (Figure 16).

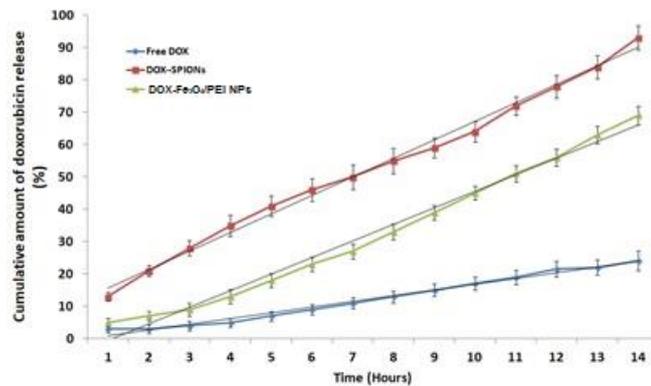

**Figure 11.** DOX release profiles of the free and the conjugated drug.

### Cellular uptake

The conjugation of PEI is expected to render the formed $Fe_3O_4$/PEI NPs with target specificity to breast cancer cells. We next quantitatively evaluated the cellular uptake of the $Fe_3O_4$/PEI NPs using ICP-OES (Fig. 6). It can be seen that after 4 h incubation with the $Fe_3O_4$/PEI NPs, MDA-MB-231 cells display significantly higher Fe uptake than MCF-10 cells, which is presumably due to the mediated specific cellular uptake of the particles. This confirms the targeting role played by PEI modified onto the surface of NPs, in agreement with the literature.

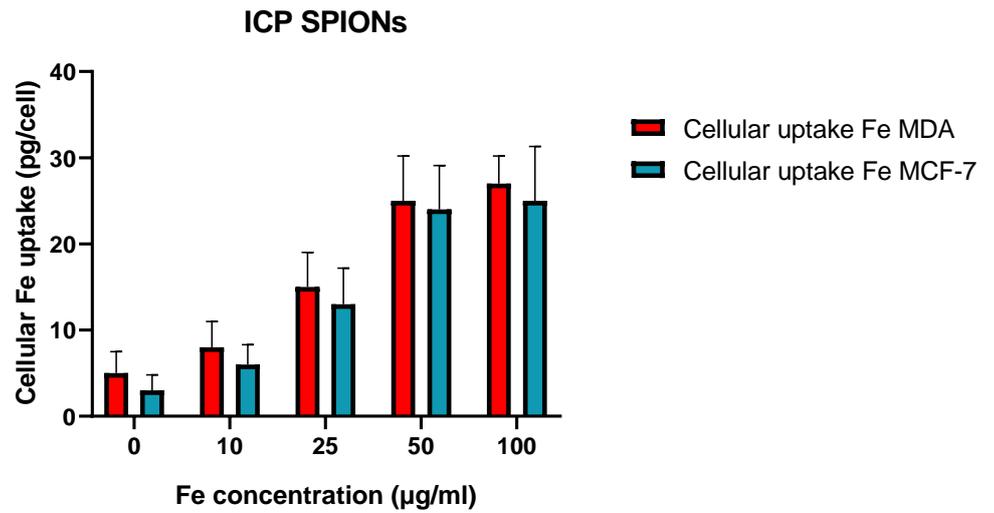

**Figure 12.** Uptake of Fe in MCF-7 or MDA-MB-231 cells treated with the Fe₃O₄/PEI NPs with different Fe concentrations (0, 10, 25, 50 and 100 µg/ml, respectively) for 4 h. MCF-7 cells treated with PBS were used as control

To further evaluate the cellular uptake of DOX-Fe₃O₄/PEI NPs in MCF-7 cells confocal microscopy studies were carried out, exploiting the intrinsic red fluorescence of doxorubicin. DOX-Fe₃O₄/PEI NPs were easily internalized in cells. It is reported the confocal microscopy images after 12 h of incubation, showing the localization of DOX-Fe₃O₄/PEI NPs in cell cytoplasm around the nucleus (Figure 13). In contrast, a lower mean fluorescence intensity was observed for the cells treated with free DOX (Figure 13).

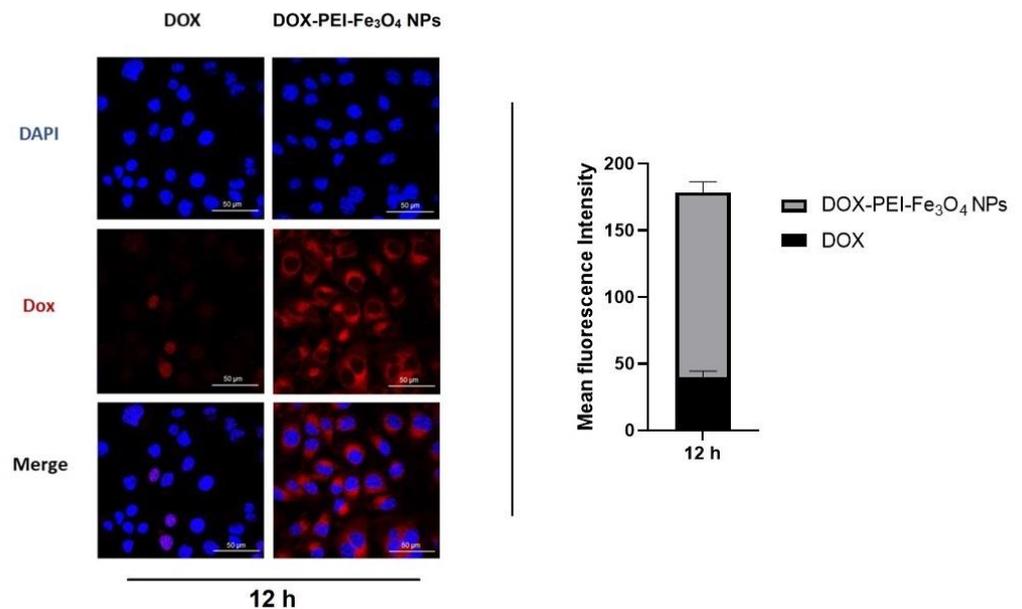

**Figure 13.** Cellular uptake studies. PEI-SPIONs-DOX and DOX were incubated with MCF-7 cells for 12 h. Exploiting doxorubicin (Dox) fluorescence, the cells were fixed and analyzed with confocal microscopy. Images are representative of three field per condition (n = 3). Quantification of Dox fluorescence inside the cells from images in panel A (n = 3).

## 4. Discussion

Superparamagnetic nanoparticles can be functionalized with doxorubicin, a very effective chemotherapeutic agent used for breast cancer therapy. Lack of specificity with inability to discriminate between malignant and noncancerous cells are one of the main causes of the side effects typical of standard DOX formulations, which limit doses and effectiveness of the drug. Moreover, DOX (or its derivatives) use is limited by cardiac and kidney toxicity and frequent onset of drug resistance in numerous tumor cell types [1]. In breast cancer cells, emergence of multidrug resistance (MDR) mainly involves upregulation of efflux transporters that are either located in the plasma or the nuclear membrane, which decrease the intracellular drug concentration [1].

Due to the well-known toxicity of Doxorubicin, new ways to deliver the drug into target tissues are urgently needed.

The magnetic core of the nanoparticles was designed to induce controlled magnetic hyperthermia by an external magnetic field. Superparamagnetic iron oxide MNPs were successfully synthesized by co-precipitation method that is facile and versatile ease of synthesizing. The great advantage of this synthesis method is high yield in the production of nanoparticles in a short time and target multidrug resistant breast cancer cells. This approach is useful to enlarge the targeting effect on these cells.

The pursuit of innovative, multifunctional, more efficient, and safer treatments is a major challenge in preclinical nanoparticle-mediated thermotherapeutic research for breast cancer. Here, iron oxide nanoparticles have the dual capacity to act as both magnetic and drug delivery agents.

DOX is widely used as a chemotherapy agent for the treatment of breast cancer [1]. Its cytotoxic and antiproliferative effects are exerted through several mechanisms, but the best known is in the poisoning of topoisomerases II cleavage complexes [2]. Doxorubicin-loaded polyethylenimine-coated superparamagnetic iron oxide nanoparticles ($Fe_3O_4$; PEI-SPIONs) with an average particle size of 30 nm, with high encapsulation efficiencies, were obtained by the electrostatic loading of doxorubicin (DOX) to SPIONs. Iron oxide-based magnetic nanoassemblies are a family of drug delivery systems with high potential like theranostic due to their low toxicity, size-dependent superparamagnetism, large surface area, and biocompatibility. In this study, we tested a novel DOX nanocarrier and its effects on normal and breast cancer cell lines.

In order to verify the targeting efficacy of nanoparticles, cell viability evaluation and 3D spheroid cell cultures cytotoxicity were conducted to estimate the viability of MCF-7 and MDA-MB-231 cells related to the in vitro efficacy of DOX, DOX-PEI-SPIONs, and DOX-PEI-SPIONs MHT with DOX concentration of 5 μM for 24, 48 and 72 hours. A gradually time-dependent decreasing trend in the cell viability with all treatments was observed. We found that the cell viability of DOX-SPIONs-treated cells was lower than that of only DOX-treated cells, and the difference was statistically significant; this may be explained by the action of DOX-SPIONs conjugates which can be easily translocated across the plasma membrane by endocytosis into breast cancer cells to exert their inhibitory effect on the cells. The viability of the DOX-SPIONs-treated cells was lower than that of the other two groups ($p<0.05$). The active targeting of conjugated nanoparticles led to more successful transport through the cellular membrane. The half inhibitory concentration of DOX-SPIONs at 24 hours was ~4.0 and ~6.4 mM for MCF-7 and MDA-MB-231 cells, respectively. Additionally, DOX released from the DOX-SPIONs was measured through dialysis at range of pH between 1.5 and 7.0 (Figure 3). Under conditions including a neutral dialysate (pH =7.4), DOX-SPIONs displayed a slow and sustained DOX release rate of ~50% within 7 hours, with ~90% of DOX released within 14 hours. These results show that DOX-SPIONs ensure a smooth and continuous diffusion of DOX in a neutral environment.

The comparison of the results of this study according to literature showed a promising ability of doxorubicin-conjugated iron-oxide nanoparticles to promote targeted apoptosis of cancer cells and a potential to act as an antimetastatic chemothermotherapeutic agent [11].

This work aimed to implement an innovative multi-therapeutic strategy that combines the use of superparamagnetic iron oxide nanoparticles properties for dual tumor targeting therapy on cancer cells by exploiting magnetic induced hyperthermia and selective drug delivery to target cancer. In our research, DOX-$Fe_3O_4$/PEI NPs showed greater targeted cytotoxicity than free DOX on breast cancer cells, regardless of DOX concentration, and lower cytotoxicity effect on mammary cell lines. Viability was decreased with increasing DOX concentration. High cytotoxicity is based on successful delivery of DOX into the nuclei of cancer cells. The treatment with DOX alone produced a very substantial decrease in cell viability. The incorporation of DOX into SPIONs core strongly enhanced the cytotoxic effect of drug when compared with the free drug.

## 5. Conclusions

DOX-$Fe_3O_4$/PEI nanoparticles proved to be a promising effective nanoformulation in the treatment of breast cancer. The increased effectiveness, compared to DOX, may be due to a higher accumulation into the tumor mass and the neoplastic cells. Moreover, the reduced biodistribution in the heart tissue suggests a reduced cardiotoxicity of DOX-SPIONs. SPIONs are promising therapeutic agents for magnetic hyperthermia of breast cancer cells and offers a dual magnetotherapeutic approach with iron oxide nanoparticles as the sole heat mediators. It was demonstrated that iron oxide nanoparticles can be remotely activated with an alternating magnetic field, achieving a very efficient heat conversion. Remarkably, the dual magnetic and drug delivery action resulted in complete cell death in vitro at low iron doses, tolerable magnetic field and frequency conditions. This cumulative, if not synergistic, heat therapy is thus promising for tumor treatment with minimal collateral tissue damage.

Therefore, this nanoformulation, which increases DOX efficacy and decreases its severe side effects, might be an attractive tool for improving the clinical management of breast cancer.

**Funding:** The research leading to these results has received funding from the European Union Seventh Framework Programme (FP7-PEOPLE-2013-COFUND) under grant agreement n° 609020 - Scientia Fellows.